\documentclass[preprint,12pt]{mycls}

\usepackage{amsmath}
\usepackage{amssymb}
\usepackage{color}

\begin{document}

\begin{frontmatter}

\title{Fractional quantum Hall states as an Abelian group} % Article title

\author{Ali Nassar}
 \ead{anassar@zewailcity.edu.eg}

\address{Center for Fundamental Physics, Zewail City of Science and Technology,
 12588 Giza, Egypt}

\begin{abstract}

We show that the set of double-layer  Fractional Quantum Hall (FQH) states with a given topological order form a finite Abelian group under a new product.  This group structure makes it possible to construct new FQH states from known ones. We also introduce a new index which can be used to characterize the topological order of FQH states.

\end{abstract}

\end{frontmatter}

Fractional quantum Hall (FQH) states \cite{Tsui:1982yy,Laughlin:1983fy} (see \cite{Wen:2004ym} for an excellent introduction)  represent novel phases of matter which arise from  the interactions of electrons in  2D layers under the influence of  strong magnetic fields ($\sim 30$ Tesla) and at very low temperatures ($\sim 1 K^\circ$). Since at very low temperature the interactions of electrons are strong, the FQH state is an example of a strongly correlated system. The new phases are not characterized by any continuous symmetry or a local order parameter as in the the usual Landau theory of symmetry breaking \cite{Wen:1989zg,Wen:1989iv,Wen:1989mee}. In \cite{Wen:1989zg,Wen:1989iv,Wen:1989mee}, it was proposed that FQH states are characterized by a new kind of order,  \textit{topological order}. This order is not the result of local long range interactions between the elementary excitations but arise from the global dancing pattern of the electrons \cite{Wenonline}.
 %but arises from the long range quantum entanglement (see \cite{Wen:2012hm} and references therein).
  One measure of topological order is the number of ground states of the Hamiltonin \cite{Haldane:1985eda}. This ground state degeneracy appears when one studies the FQH state on a genius-$g$ Riemman surface, e.g., on a two torus which  is equivalent to a layer with periodic boundary conditions in the two directions. The ground state degeneracy is a universal property and is robust against local perturbations \cite{Wen:1990zza}.

The kind of order one gets in FQH states is different from crystal order. Crystal order is static and arises from the spatial configuration and location of the atoms inside the crystal. {{In this note, we study the properties of the ground state of a FQH system described by an effective Chern-Simons theory. The ground state in this case is not specified by the values of a set of local order parameters and it describes a topological phase.
The order in topological phases is dynamical and arises from the dancing patterns of the electrons in FQH states. This dancing pattern is specified by an integer symmetric matrix  $K$ \cite{Wen:1992uk}.
 In this note, we define an Abelian group structure on the set of FQH states with a given topological order defined in terms of the determinant of $K$.  The topological order characterizes the universality classes of FQH states in a given topological phase and our result shows that these states enjoy an Abelian symmetry group. This Abelian symmetry group can give rise to a set of discrete symmetries or quantum numbers which could be useful in characterizing the topological phase. }}

% $SL(2,\mathbb{Z})$ equivalence classes of $K$ with a given determinant.

%The CFT description \cite{Wen:1994ha}

%\section*{Effective theory of FQH states}

We will be interested in Abelian  FQH states, where Abelian here means that the quasi-particle excitations follow Abelian statistics.
The effective theory of such states is a $U(1)$  Chern-Simons gauge theory in $2+1$ dimensions described by the action \cite{Wen:1992uk}
\begin{equation}\label{cs1}
S=-\frac{k}{4\pi}\int_M A \wedge d A+\frac{q}{2\pi}  a \wedge dA,
\end{equation}
where $k\in\mathbb{Z}$ is the level of the Chern-Simons term and $A_\mu$ is the gauge field which is dual to the quasiparticle conserved current, $J_\mu = \epsilon_{\mu\nu\alpha}\partial_\nu  A_\alpha $. Here  $a_\mu$ is an external $U(1)$ gauge field which couples to a quasiparticle excitations with charge $q$. One can also add a Yang-Mills term to the above action but the Yang-Mills term is an irrelevant operator in 3D and at low energies the Yang-Mills term is dominated by the Chern-Simons term. {{An example of a physical system which is described by the effective action (\ref{cs1}) is a 2D electron gas in the first Landau level.}}

{{A more general situation happens when the electron density is such that $N$ Landau levels are filled}. If the energy gap between the different levels is large enough then we can treat the different levels as dynamically independent \cite{Zee:1996fe}. In this case, we can take the current belonging to each level as being separately conserved
\begin{equation}
d J^I =0, \qquad J^I=d A^I,   \quad I=1,\cdots,N.
%\partial^\mu J_{\mu}^I=0,\qquad J_{\mu}^I = \epsilon_{\mu\nu\alpha}\partial_\nu  A_\alpha^I
\end{equation}
The free action of the system is
\begin{equation}
S=-\frac{k_I}{4\pi}\int_M A^I \wedge d A^I, \quad I=1,\ldots,N.
\end{equation}
One can couple the different levels to one another through electron interactions which in turn  changes the low energy effective action to
\begin{equation}
S=-\frac{k_I}{4\pi}\int_M A^I \wedge d A^I+\frac{\tilde{K}_{IJ}}{4\pi} A^I\wedge J^J, \quad I=1,\ldots,N.
\end{equation}
Using $J^J=dA^J$, the above action can be written succinctly as
\begin{equation}\label{cs}
S=-\frac{1}{4\pi}K_{IJ}\int_M A^{I} \wedge d A^J+\frac{1}{2\pi} q_I a \wedge dA_I, \quad I=1,\ldots,N,
\end{equation}
where $K_{IJ}$ is a symmetric integer matrix and we also added a coupling to an external gauge field $a$.
}

The action (\ref{cs}) of the the multi-layer FQH states is  an $U(1)^N$ Chern-Simons theory where index $I$ characterizes the different condensates of the FQH state \cite{Wen:2004ym}. In the canonical quantization of (\ref{cs})  one first divide the three manifold $M$ as $M=\mathbb{R}\times \Sigma_g$, where $\mathbb{R}$ represents the time direction and  $\Sigma_g$ is the genus-$g$ Riemann surface which represents the two-dimensional layer over which the FQH state lives. The quantum mechanical degrees of freedom
are given by the  harmonic parts of $A_I$ on $\Sigma_g$ and correspond to non-trivial global configurations of $A_I $.
{{Here we briefly review the quantization of the above action following \cite{KeskiVakkuri:1993xp}. To proceed with the quantization, we add to the action a Yang-Mills term
\begin{equation}
S=-\frac{1}{4\pi}K_{IJ}\int_M A^{I} \wedge d A^J+ \frac{1}{g_{\text{YM}}}F_I\wedge *F_I ,
\end{equation}
where $g_{\text{YM}}$ is the Yang-Mills coupling which has the dimension of mass in 3D. The final results are independent of $g_{\text{YM}}$ and the Yang-Mills term acts as an IR regulator and one can take $g_{\text{YM}}$ to infinity at the end. By working in Weyl  gauge $A_{I0}=0$  then one can write the Hodge decomposition of $A_{Ii}$ on $\Sigma_g$ as
\begin{equation}
A_{Ii}=\theta_{Ii} + a_{Ii},
\end{equation}
where $a_{Ii}$ is the pure gauge piece and $\theta_{Ii}$ is the global part of $A_I$.

The variation of the action with respect to $A_{I0}=0$ leads to the constraint
\begin{equation}
0=\frac{\delta S}{\delta A_{I0}}=-\frac{1}{4\pi} K_{IJ} F_{Jij} +\frac{1}{g_{\text{YM}}} h^{ij} \partial_i F_{I 0j}
\end{equation}
where $h^{ij}$ is the metric of $\Sigma_g$. Using this in the action one can write
\begin{equation}
S= \int dt \bigg[\frac{K_{IJ}}{4\pi}\big(\theta_{I2} \dot{\theta}_{J1} -\theta_{I1} \dot{\theta}_{J2}\big)+ \frac{1}{2} m^{ij} \dot{\theta}_{Ii} \dot{\theta}_{Ij}\bigg] +\tilde{S}_{\text{loc}}
\end{equation}
where $m^{ij}=\frac{1}{g_{\text{YM}}} h^{00} h^{ij}$ and $\tilde{S}_{\text{loc}}$ is independent of $\theta$.  One can easily write down the Hamiltonian from the above action
\begin{equation}
H=\frac{1}{2}(m^{-1})_{ij}\sum_I \bigg(\frac{\partial}{\partial \theta_{Ii}} -A^{\theta}_{Ii} \bigg)\bigg(\frac{\partial}{\partial \theta_{Ij}} -A^{\theta}_{Ij} \bigg)
\end{equation}

The ground state wavefunction $\psi_0(\theta_{Ii})$ of the above Hamiltonian turned out to be degenerate and it was shown in \cite{KeskiVakkuri:1993xp} that the degeneracy  $D$ is given by
\begin{equation}
D=\det(K)^g.
\end{equation}
This degeneracy depends only on the topology of the FQH sample and it disappears if we put the sample on a sphere. This degeneracy is robust against any local perturbations of the sample and it represents a good measure of the topological order in the ground state.
}}

The action (\ref{cs}) is supplemented with a quantization condition on the set of allowed charges of the quasiparticles. This quantization condition implies that the set of allowed charges live in an integer lattice and the
 gauge group is the compact $U(1)^N$. Due to this quantization condition, the allowed field redefinitions on the set of $A^I$ are those transformations which preserve
the integrality of the charges, i.e, $SL(N,\mathbb{Z})$ transformation
\begin{equation}\label{slnz}
A_I\longrightarrow S_{IJ} A_J, \quad S \in SL(N,\mathbb{Z}).
\end{equation}

In \cite{Halperin:1983zz}, the multi-layer FQH stat which is described by (\ref{cs}) was proposed
\begin{equation}\label{dlfqhs}
\Psi_K(z)=\prod_{a<b;I,J}(z_{aI}-z_{bJ})^{K_{IJ}} \exp\Big(-\sum_{a,I} |z_{aI}|^2\Big).
\end{equation}
This state is labeled by the matrix $K$  and a charge vector $\mathbf{t}$ and describes a FQH state with a filling fraction\footnote{The filling fraction is the {{ratio}} of the electron density to the flux density.}
\begin{equation}\label{fillingfraction}
\nu=\mathbf{t}^T K^{-1} \mathbf{t}.
\end{equation}
The above state is a generalization of the one-component Laughlin state
\begin{equation}
\Psi_m(z)=\prod_{a<b}(z_{a}-z_{b})^m \exp\Big(-\sum_a |z_{a}|^2\Big).
\end{equation}

The $SL(N,\mathbb{Z})$ transformation (\ref{slnz}) on $A_I$  takes the following form on $\mathbf{t}$ and $K$
\begin{equation}
\mathbf{t}\longrightarrow S \mathbf{t}, \qquad K\longrightarrow S K S^T, \quad S \in SL(N,\mathbb{Z}).
\end{equation}
This transformation leaves the filling fraction invariant. {{Using the invariance of the action (\ref{cs})} under the $SL(N,\mathbb{Z})$ transformation (\ref{slnz}), then the state (\ref{dlfqhs}) is effectively characterized by the $SL(N,\mathbb{Z})$ equivalence classes of $K$ where the equivalence relation is
\begin{equation}\label{ss}
K\sim K', \quad \text{iff } K'=S K S^T,\qquad S\in SL(N,\mathbb{Z}).
\end{equation}}

%The ground state degeneracy of  Abelian FQHS on the torus is given by $D=\det(K)$. This degeneracy is not a consequence of a %local symmetry of the Hamiltonian and is purely topological in nature.

One of the quantum numbers which can be used to characterize the  topological order of the state (\ref{dlfqhs}) is the ground state degeneracy on the torus \cite{Wen:1989iv,KeskiVakkuri:1993xp}
\begin{equation}\label{toporder}
D=\det{K}.
\end{equation}

Here we limit ourselves to the double-layer FQH states, where in this case the matrix $K$ is a $2\times 2$ matrix subject to the $SL(2,\mathbb{Z})$ transformation (\ref{slnz}). The topological order (\ref{toporder}) is an invariant of the $K$ matrix under (\ref{slnz}), i.e, it's the same for all members of the $SL(2,\mathbb{Z})$ equivalence class.
Since $S=\pm I$ in (\ref{ss}) acts trivially, then only the group $PSL(2,\mathbb{Z})$ acts effectively.
Now we can talk of the equivalence classes of $K$ with $\det(K)=D$  under the action of $PSL(2,\mathbb{Z})$. We simply take all the matrices which are in the same orbit of $K$ under $PSL(2,\mathbb{Z})$ as one equivalence class.
The set of  $PSL(2,\mathbb{Z})$ classes is defined as \cite{Hosono:2002yb}
\begin{equation}\label{ClD}
 Cl(D) =
  \bigg\{
  K=\begin{pmatrix}a&b\\ b&c \end{pmatrix} \, \big|\, D=ac-b^2 >0,\; a>0  \bigg\} / \sim SL(2,\mathbb{Z}).
\end{equation}
We assume that the greatest common divisor  $\text{gcd}(a,b,c)=1$ which restricts us to primitive equivalence classes.

The set $Cl(D)$ is finite and its cardinality will be denoted by $h(D)=|Cl(D)|$. The elements in $Cl(D)$ will be written as
\begin{equation}
Cl(D)=\{\mathcal{C}_1,\cdots,\mathcal{C}_{h(D)}\}.
\end{equation}
We will denote the matrix $K$ by $K(a,b,c)$ and the equivalence class of $K$ under $PSL(2,\mathbb{Z})$ will be denoted by $\mathcal{C}=[K(a,b,c)]$.

%It is know that
%The double layer FQHS with topological order given by a symmetric matrix $K$ with integer entries and charge vector $Q$ will have %a filling fraction given by
%\begin{equation}
%\nu = Q^T K^{-1} Q,
%\end{equation}
%where $K$ is an understood to be an $SL(2,\mathbb{Z})$ equivalence class. The topological order of this state is $D=\det(K)$ and %is the same for all the members of the $SL(2,\mathbb{Z})$ equivalence class. Now, another state with a different $SL(2,\mathbb{Z}%)$ equivalence class $K'$ and same topological order satisfies $D=\det(K')$ and will have a different filling fraction. The Gauss %product will be used to define an Abelian group structure on those states.

%We can associate to $Q$ a $2\times 2$ matrix
%\begin{equation}
%M(Q)=\begin{pmatrix}
%2a &b\\
%b&2c
%\end{pmatrix}.
%\end{equation}
%The discriminant of $Q$ is now given by  $D=-\det(M)$.

%The matrix representation allows us to define an equivalence relation.

%\subsection*{The Gauss product}

There is a binary operation which can be defined on the set $Cl(D)$ and it turns $Cl(D)$ into a finite Abelian group.  This operation is called the Gauss product (see \cite{Hosono:2002yb}) and it takes two equivalence classes in $Cl(D)$ and produces a third equivalence class with same determinant.

Let  $\mathcal{C}_1=[K_1(a_1,b_1,c_1)]$ and $\mathcal{C}_2=[K_2(a_2,b_2,c_2)]$ be two such equivalence classes.  We say that two quadratic forms $K_1(a_1,b_1,c_1)\in \mathcal{C}_1$ and $K_2(a_2,b_2,c_2)\in \mathcal{C}_2$ are concordant if $a_1 a_2 \neq 0$,  $\text{gcd}(a_1,a_2)=1$ and $b_1=b_2$. Then the Gauss product of  $\mathcal{C}_1\star\mathcal{C}_2=\mathcal{C}_3$ is defined as
\begin{equation}
\big[K_1\big(a_1,b,c_1\big)\big] \star\big[K_2\big(a_2,b,c_2\big)\big]=\big[K_3\big(a_3,b_3,c_3\big)\big],
\end{equation}
where $a_3=a_1 a_2$, $b_3=b$, and $c_3=\frac{b^2+D}{4a_1 a_2}$.
It is important to mention that any pair of quadratic forms of the same discriminant can be $SL(2,\mathbb{Z})$-transformed to a concordant pair.

The identity element  $\mathbf{1}_D$ of $Cl(D)$ with respect to the product $\star$ is represented by
\begin{equation}
\mathbf{1}_D=
\begin{cases}
[1,0,\frac{D}{4}]; & \text{if } D\equiv 0 \mod 4\\[5pt]
[1,0,\frac{1+D}{4}]; & \text{if } D\equiv 1 \mod 4.
\end{cases}
\end{equation}

Now that we defined the Gauss product, we can use it to show that the set of double-layer FQH states (\ref{dlfqhs}) is a finite Abelian group $\mathcal{G}$ of order $h(D)$.
As we explained before, the
double-layer FQH states are characterized by an $SL(2,\mathbb{Z})$ equivalence class  $[K]$. We define the following product on the set of double-layer FQH states:
\begin{equation}\label{new-product}
\Psi_{[K_1]} (z) \odot\Psi_{[K_2]} (z) =\Psi_{[K_3]} (z),
\end{equation}
 where $K_3$ is given by the Gauss product of $[K_1]$ and $[K_2]$
 \begin{equation}
 [K_3]=[K_1]\star[ K_2].
 \end{equation}
 The group property of $\odot$ follows directly from the group property of $\star$. As far as we know, this kind of group structure on the set of FQH states didn't appear before in the study of the FQHE. The dimension of $\mathcal{G}$ is equal to the number of equivalence classes of $K$ with determinant $D=\det(K)$ which is nothing but the  number $h(D)$. As we have seen in (\ref{toporder}), $D$ gives the degeneracy of the ground state of the FQH systems which is a measure of topological order \cite{Wen:1989zg,Wen:1989iv,Wen:1991rp}.  We propose the number $h(D)$ as another index which characterizes the topological order of the double-layer FQH states. One can use the group property of $\mathcal{G}$ to generate new FQH states (with the same topological order and different filling fractions) from older ones. For example, a new state $\Psi_{[K_3]} (z)$ can be constructed out of two known states $\Psi_{[K_1]} (z)$  and  $\Psi_{[K_2]} (z)$ by simply multiplying them using $\odot$.   Wether these new states have been constructed before or realized experimentally is unknown to the author.

We recall the
\textit{fundamental theorem of finite Abelian groups} which states that that every finite Abelian group is isomorphic to a direct product of cyclic groups of prime-power order, where the decomposition is unique up to the order in which the factors are written. Using this fact, we can write the order-$h(D)$ Abelian group $\mathcal{G}$ as
\begin{equation}
\mathcal{G}=\mathbb{Z}_{p_1^{n_n}}\times\cdots\times \mathbb{Z}_{p_n^{n_n}},
\end{equation}
where
\begin{equation}
h(D)=p_1^{n_n}\times\cdots\times p_n^{n_n}.
\end{equation}
This implies that the set of FQH states enjoys a set of discrete symmetries which emerges entirely from the dynamics. This is not a symmetry of the underlying Hamiltonian but a property of the $K$-matrix characterization of the FQH states. Since $K$ is related to the dynamical motion of the electron, the symmetry $\mathcal{G}$ is dynamical. {{This new symmetry could be of physical interest since it leads to a new set of discrete quantum numbers which could be used to characterize the universality classes of topological phases with a given topological order.}}

I am grateful to Mark Walton for useful discussions. This research is supported by the CFP at Zewail city of Science and Technology. {We would like to thank the anonymous referee for useful comments.}

\bibliographystyle{elsarticle-num}
\bibliography{refs}

%\end{multicols}
\end{document}